\documentclass[pra,twocolumn,showpacs,floatfix,amsmath,amsfonts]{revtex4}
\usepackage{amsmath}
\usepackage{epsfig}
\usepackage{amssymb}
\begin{document}

\title{Symmetry-assisted vorticity control in Bose-Einstein condensates}

\author{V\'{\i}ctor M. P\'erez-Garc\'{\i}a}
\affiliation{Departamento de Matem\'aticas, Escuela T\'ecnica
Superior de Ingenieros Industriales, \\
Universidad de Castilla-La Mancha, 13071 Ciudad Real, Spain}

\author{ Miguel A. Garc\'{\i}a-March}
\affiliation{Departamento de Matem\'aticas, Escuela T\'ecnica
Superior de Ingenieros Industriales, \\
Universidad de Castilla-La Mancha, 13071 Ciudad Real, Spain}

\author{Albert Ferrando}
\affiliation{Interdisciplinary Modeling Group, InterTech. Departament d'\`{O}ptica,
Universitat de Val\`{e}ncia. Dr. Moliner, 50. E-46100 Burjassot (Val\`{e}ncia),
Spain.}

\begin{abstract}
Using group-theoretical methods and numerical simulations we show how to act on the topological charge of individual vortices in Bose-Einstein condensates by
using control potentials with appropriate discrete symmetries. As examples of our methodology we study charge inversion and  vortex
erasing by acting on a set of control laser gaussian beams generating optical dipole traps.
\end{abstract}


 \maketitle

\section{Introduction}

Vortices have been a source of fascination
since the works of Empedocles, Aristotle and Descartes, who tried
to explain the formation of the Earth, its gravity and the
dynamics of the solar system as due to primordial cosmic vortices.
Many interesting problems related to vortices are open in different
fields such as fluid mechanics, superconductivity, superfluidity,
light propagation, Bose-Einstein condensation (BEC), cosmology,
biosciences, or solid state physics \cite{Lug95,Pis99,Sols,experimental1,experimental2,experimental3}.

In wave mechanics a vortex is a screw phase
dislocation, or defect \cite{nye74}, where the amplitude of the
field vanishes. The phase around the singularity has an integer
number of windings, $\ell$, which plays the role of an angular
momentum. For symmetric systems, this
number is a conserved quantity and governs the interactions
between vortices as if they were endowed with electrostatic
charges. Thus, $\ell$ is usually called the \emph{topological
charge} of the defect.

Angular momentum is conserved in a quantum system
with O(2) rotational symmetry. If we consider a
state with well-defined angular momentum $ \ell \in \mathbb{Z}$, i.e.,
an eigenfunction of the angular momentum operator at a given time $t_0$, its
evolution will preserve the value of $\ell$. In a system possessing a discrete
point-symmetry (described by the $C_n$
and $C_{nv}$ groups) the angular momentum is no longer
conserved. However, in this case one can define another
quantity $m \in \mathbb{Z}$, the Bloch or angular pseudo-momentum,
which is conserved under time evolution \cite{9}. The angular
pseudo-momentum $m$ plays then the role of $\ell$ in a system with
discrete rotational symmetry. From the group theoretical point
of view, the angular momenta and pseudo-momenta, $\ell$ and $m$,
are also the indices of the 2D irreducible representations of
O(2) and $C_n$, respectively \cite{10,11,12}. Unlike $\ell$, the values of
$m$ are limited by the order of the point-symmetry group $C_n$.

The existence of vortices is one of the signatures of superfluidity and this is why in the field of Bose-Einstein
condensation, they have attracted so much interest. They can be generated in rotating traps  \cite{experimental2,experimental3} or by phase imprinting methods \cite{Kett1,Kett2}. The later procedure allows to generate only multiply charged vortices with topological charges $m=2,4$.

In the last years there has been an enormous  interest on the applications of group theory to study the
properties of defects in media with discrete symmetries including  photonic crystals,  periodic potentials
with discrete symmetries, etc \cite{9,10,11,12,SoliTop,Oster}.

In this paper we explore the application of group theory  to
control the topological charge of vortices in Bose-Einstein condensates by using external potentials
with discrete rotational symmetry.  To do so we propose a simple setup based on a non-periodic potential with discrete rotational symmetry which will allow us to perform many operations with the vortex charges depending on the initial charge and the potential symmetry order. While the vortex transmutation has been previously explored in potentials with broken symmetries \cite{Ripi} and in the context of photonic lattices \cite{13}, other operations to be proposed here have not been studied before.
Our proposal is simpler to implement than the 2D lattice type potentials proposed in the framework
of photonic lattices \cite{SoliTop} and is easier to reconfigure.

 We will also show how starting from multiply charged vortices such as the ones which can be generated in  atom chips by phase-imprinting  methods \cite{Kett1,Kett2} one can generate different types of vortices by choosing an appropriate control potential.

Our plan is as follows: In Sec. \ref{III} we present the theory in which our methodology is based. In Sec. \ref{III} we present several examples. First we present our specific setup in Sec. \ref{Pra} and then discuss several phenomena which can be achieved such as: topological charge erasing (Sec. \ref{erase}), and two examples of topological charge inversion from $m=2$ to $m=-1$ (Sec. \ref{single}) and $m=4$ to $m=-1$ (Sec. \ref{minusingle}).

\section{Theory}
\label{III}

In this paper we will consider a Bose-Einstein condensate with tight confinement along an specific direction ($z$) leading to a quasi-two dimensional Bose-Einstein condensate.
Under the effect of an additional external potential $V(\mathbf{x})$ this system
 is ruled in the mean field limit  by an effective Gross-Pitaevskii equation \cite{PG98,Ripi}
\begin{equation}\label{NLS}
i \partial_t \psi(\mathbf{x}) = -\frac{1}{2} \Delta_{\mathbf{x}} \psi(\mathbf{x}) + V(\mathbf{x}) \psi(\mathbf{x}) + g |\psi(\mathbf{x})|^2 \psi(\mathbf{x}),
\end{equation}
in dimensionless units and where $g$ is a measure of the effective nonlinearity.

The symmetry of the potential induces strict restrictions on the vorticity of stationary solutions of Eq.(\ref{NLS}). In order to illustrate this statement, let us consider the equation for stationary solutions, which is a nonlinear eigenvalue equation of the
following type:
\begin{equation}\label{SNLS}
H \left(\psi(\mathbf{x}),\mathbf{x} \right) \psi(\mathbf{x})=\mu \psi(\mathbf{x}),
\end{equation}
where the nonlinear Hamiltonian operator depends on the field itself, i.e., $H\left(\psi(\mathbf{x}),\mathbf{x} \right) \equiv -\frac{1}{2} \Delta_{\mathbf{x}}  + V(\mathbf{x})  + g |\psi(\mathbf{x})|^2$  The nonlinear solution $\psi$ can be considered as a self-consistent solution. It appears as an eigenstate of the Hamiltonian but, at the same time, it defines the Hamiltonian operator itself since $H$ depends on it through the nonlinear term $g |\psi|^2$.

\begin{figure}
\epsfig{file=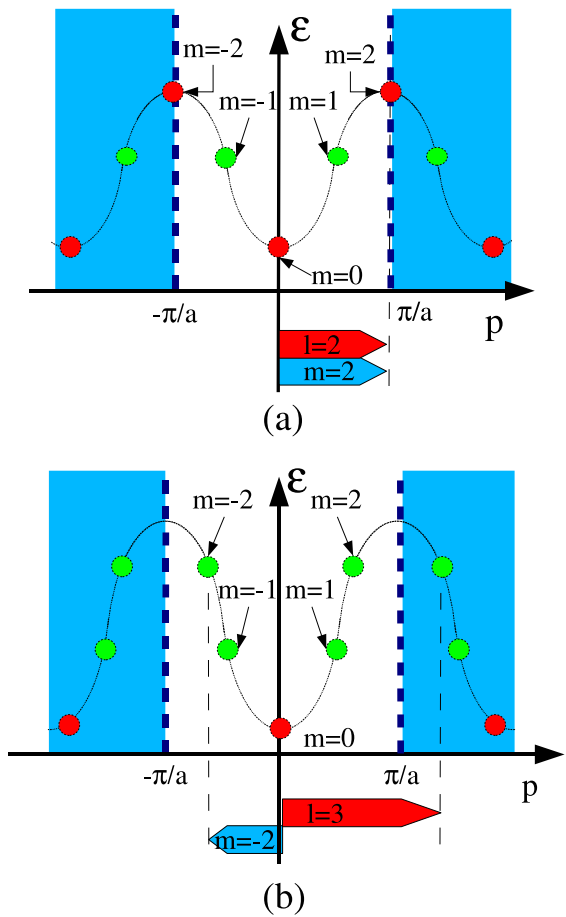,width=\columnwidth}
\caption{[Color online] Angular Bloch structure of solutions for a system with rotational symmetry of order: (a) $N=4$ and (b) $N=5$.  Red and green circles symbolize solutions that belong either to one dimensional or two dimensional irreducible representations.  Red arrows represent input angular momentum while blue arrows represent output angular pseudo-momentum. Some examples of matching conditions for angular pseudo-momentum are indicated.\label{Bloch_structure}}
\end{figure}

One can establish necessary conditions for the existence of stationary symmetric solutions of Eq.(\ref{SNLS}) based on the symmetry properties of the potential $V(\mathbf{x})$.  They will exhibit special features due to the nonlinear nature of the Hamiltonian operator $H(\psi)$. We can summarize the properties of these solutions in a single statement: (i) if the potential $V$ is invariant under a point symmetry group that we refer to generically as ${\cal G}$ ---i.e., ${\cal G}$ describes finite two-dimensional $2 \pi/N$ rotations around an axis ($C_N$ group) and specular reflections  ($C_{Nv}$ group)--- and (ii) if we search for symmetric solutions fulfilling the condition $|\psi(G\mathbf{x})|^2=|\psi(\mathbf{x})|^2$ (where $G$ is any element of the group ${\cal G}$), then the solution $\psi$
must belong to some representation $D^m_\nu({\cal G})$ of the symmetry group ${\cal G}$ or to some of their subgroups ${\cal G'}\subset {\cal G}$. Since this statement is rather mathematical, it is convenient to analyze it in the light of the symmetry properties of the nonlinear Hamiltonian $H(\psi)$.  First of all, let us recall that if $\psi$ is a given stationary solution  $\psi=\psi_{\textrm{sol}}$ satisfying Eq.(\ref{SNLS}) then $\psi_{\textrm{sol}}$ plays two different roles in this nonlinear eigenvalue equation. On the one hand, $\psi_{\textrm{sol}}$ defines the Hamiltonian operator $H(\psi_{\textrm{sol}})$ whereas, on the other hand, it appears as an eigenfunction of the same operator. This two-fold role has profound implications on the allowed functional form of the solution. The first consequence of the explicit, and specific, dependence of the nonlinear Hamiltonian on $\psi_{\textrm{sol}}$ is that $H(\psi_{\textrm{sol}})$ inherits the symmetry of the potential. This is a consequence of assumptions (i) and (ii). Since the Laplacian operator is invariant under any type of rotation $\Delta_{G\mathbf{x}}=\Delta_{\mathbf{x}}$ and, on the other hand, $V(G\mathbf{x})=V(\mathbf{x})$ and $|\psi_{\textrm{sol}}(G\mathbf{x})|^2=|\psi_{\textrm{sol}}(\mathbf{x})|^2$ because of the previous assumptions, the nonlinear Hamiltonian evaluated at $\psi{_\textrm{sol}}$ is automatically invariant under the symmetry group $\cal{G}$:
$H\left( \psi(G \mathbf{x}), G \mathbf{x} \right) =-\frac{1}{2}\Delta_{G\mathbf{x}}+V(G\mathbf{}x) + g |\psi(G \mathbf{x})|^2=-\frac{1}{2} \Delta_{\mathbf{x}}  + V(\mathbf{x})  + g |\psi(\mathbf{x})|^2=H\left(\psi(\mathbf{x}),\mathbf{x} \right)$.  The fact that the Hamiltonian operator evaluated at $\psi_{\textrm{sol}}$ is invariant under the group $\cal{G}$ implies that $H(\psi_{\textrm{sol}})$ commutes with all elements of this group, i.e., $[H(\psi_{\textrm{sol}}),G]=0$ $\forall G \in {\cal G}$ and, therefore, according to standard quantum mechanics arguments, all its eigenfunctions must belong to the different representations of the symmetry group ${\cal G}$. At this point, we have considered $\psi_{\textrm{sol}}$ only in its role of generator of the nonlinear Hamiltonian operator. However, in its second role $\psi_{\textrm{sol}}$ must also appear as an eigenfunction of $H(\psi_{\textrm{sol}})$. Therefore, $\psi_{\textrm{sol}}$ must belong to some of the representations of $H(\psi_{\textrm{sol}})$, i.e., $\psi_{\textrm{sol}}\in D^m_\nu({\cal G})$, where $D^m_\nu$ indicates the representation characterized by the representation index $m$ and the degeneracy index $\nu$. The degeneracy of the representations of the point symmetry groups $C_N$ and $C_{Nv}$ is either one or two \cite{11}. One dimensional representations automatically fulfill the symmetry condition for the amplitude $|\psi(G\mathbf{x})|^2=|\psi(\mathbf{x})|^2$ assumed in (ii)  since they do not transform, except for a sign, under the action of a finite rotation: $G \psi(\mathbf{x})=\psi(G\mathbf{x})=\pm \psi(\mathbf{x})$. The requirement of amplitude invariance is, however, trickier for two-dimensional representations. Let us see next why. A natural basis for a two-dimensional representation of a point symmetry group is that formed by the eigenvectors of the group operator in such representation, i.e., that given by the two functions $(\psi_m, \psi^\ast_m)$ fulfilling $G \psi_m(\mathbf{x})=\psi_m ({G\mathbf{x})=\epsilon^m \psi_m(\mathbf{x})}$ and its complex conjugate, $m$ being the representation index and the eigenvalue being given by $\epsilon=e^{i 2 \pi/N}$. Since a group rotation operator acting on this representation will be represented by a $2 \times 2$ matrix this two vectors provides the basis where this matrix is diagonal. However, the more general form of a function belonging to the $m$-representation will be given by a linear combination of $\psi_m$ and $\psi^\ast_m$. This general solution has a problem with respect to the requirement of invariance for the amplitude assumed in (ii). Indeed, in the most general case in which we consider arbitrary coefficients, this linear combination does not verify the previous condition and thus such a function would be excluded as a solution of the problem enjoying symmetry under the full group ${\cal G}$ (however, it can fulfill the condition for a subgroup ${\cal G'} \subset {\cal G}$, see Ref.\cite{10}). In this two-dimensional subspace the only two functions fulfilling the condition are the eigenvectors of the $G$ operator $\psi_m$ and $\psi^\ast_m$. This is so because the eigenvalue $\epsilon^m$ is a pure phase number and therefore with unit modulus. Consequently, only $\psi_m$ and $\psi^\ast_m$ can be nonlinear solutions of Eq.(\ref{SNLS}). This fact is in remarkable contrast with respect to the linear case in which all linear combinations belonging to the representation appear as solutions of the eigenvalue equation. Notice that the $\psi_m$ and $\psi^\ast_m$ complex solutions represent a vortex-antivortex soliton pair of topological charge $m$ and $-m$ respectively. In order to interpret this result on a more physical basis it is convenient to re-write the action of a discrete rotation of order $N$ on these functions using polar coordinates:
\begin{equation}
G \psi_m(r,\theta)=\psi_m(r,\theta+2\pi/N)=e^{i 2 \pi m/N}\psi_m(r,\theta)
\label{transformation_property}
\end{equation}
and its complex conjugate. As it was recognized in Ref.(\cite{9}), this condition is identical to that fulfilled by one-dimensional Bloch modes but by substituting a standard non-compact spatial coordinate by the compact angular one $\theta$. For this reason the most general form for such solutions are that corresponding to angular Bloch modes:
\begin{equation}\label{bloch_modes}
\psi_m(r,\theta)=e^{i m \theta}f_m(r,\theta)
\end{equation}
where $f_m$ is an invariant function under $2 \pi/N$ rotations and $m$ plays the role of ``angular" pseudo-momentum satisfying the following constraints: $|m|<N/2$ (if $N$ is even) and $|m|\le (N-1)/2$ (if $N$ is odd). Since $m$ is also a representation index, these constraints can be explained using group theory arguments \cite{12}. However, it is physically more intuitive to interpret them by means of the properties of the angular pseudo-momentum in an equivalent Bloch problem in terms of the angular variable $\theta$. The symmetry condition of the potential under $N$th-order rotations ---$V(r,\theta+2 \pi/N)=V(r,\theta)$--- is understood as a periodicity condition in the angular variable $\theta$ with period given by $a=2 \pi/N$. As we have seen, this periodicity property is inherited by the full Hamiltonian $H(\psi)$ when $\psi$ fulfills the symmetry condition $|\psi(r,\theta+2 \pi/N)|^2=|\psi(r,\theta)|^2$ (assumption (ii)). Since the full Hamiltonian is periodic it is then natural that their solutions transform as in Eq.(\ref{transformation_property}) and they present the angular Bloch form (\ref{bloch_modes}). The role of $m$ is therefore that corresponding to the conjugate variable of the periodic variable $\theta$, hence its name of ``angular" pseudo-momentum. The angular nature of $\theta$ forces $m$ to be an integer in order to preserve the single-valuedness of the wavefunction $\psi$. Energy is a periodic function of pseudo-momentum with periodicity given by the so-called reciprocal lattice vector $2\pi/a=N$, which determines the existence of equivalent zones in pseudo-momentum space called Brillouin zones. In this framework the constraints in the values of $m$ arise from the fact that all possible solutions can be reduced to those existing in the first Brillouin zone in pseudo-momentum space, i.e., those fulfilling $|m|\le \pi/a=N/2 $, which is equivalent to the aforementioned constraints. As an example, in Fig.\ref{Bloch_structure} it is represented the angular Bloch structure of energy eigenstates of two systems owning discrete symmetry of $4$th and $5$th order. Notice that the period of the Brillouin zone ---i.e., the reciprocal lattice vector--- is equal to the order of symmetry ($N=4,5$) and that the allowed angular pseudo-momenta fullfill the condition $m \le 2$ corresponding to the first Brillouin zone. In this scenario it is rather intuitive to understand what happens to a solution carrying angular momentum $l$ evolving in a medium characterized by full rotational invariance ---given by the $O(2)$ group--- when we suddenly switch on a potential that breaks full rotational symmetry into a discrete-symmetry of $N$th order ---given by the $C_N$ group. In terms of the angular variable $\theta$ the solution with angular momentum $l$ behaves like a sort of ``angular" plane wave $e^{i l \theta}f_l(r)$ since the amplitude $f_l(r)$ is angle-independent. The angular momentum $l$ plays the role of ordinary (discretized) momentum. The problem is thus equivalent to that of a plane wave propagating in a constant potential ---in $\theta$, since $U(r)=V(r)+g f^2_l(r)$ is angle independent--- that, at some specific moment $t_0$, feels the presence of a periodic potential $U(r,\theta+a)=U(r,\theta)$. At $t_0$ the wavefunction corresponding to the initial angular plane wave must excite the spectrum of the nonlineal operator $H(\psi(t_0))$. However, the full potential $U(r,\theta;t_0)=V(r,\theta)+g f^2_l(r)$ is no longer $O(2)$-invariant but $C_N$-invariant due to the breaking of continuous symmetry by the appearance of the discrete-symmetry potential $V(r,\theta)$ at $t_0$ and whose eigenstates are angular Bloch modes of the form (\ref{bloch_modes}). This fact implies that there must be a matching between the input angular momentum and the angular pseudo-momenta of allowed angular Bloch states after the $C_N$ interaction is switched on. In other words, the initial angular momentum $l$ carried by the angular plane wave $e^{i l \theta}f_l(r)$ must match the angular pseudo-momentum $m$ of some angular Bloch state. If $|l|\le N/2$ the initial angular momentum can always match an angular pseudo-momentum in the first Brillouin zone and, consequently, $l=m$. However, if $|l| > N/2$ we excite angular Bloch states in higher-order Brillouin zones which, on the other hand, we know are equivalent to those in the first Brillouin zone. This equivalence is given by periodicity in pseudo-momentum space which is provided by the reciprocal lattice vector $2\pi/a=N$ in such a way two pseudo-momenta are equivalent if they differ from each other by a multiple of this vector. This establishes the desired matching condition for angular pseudo-momentum in terms of the initial angular momentum $l$:
\begin{equation}
l-m=kN,\;k \in \mathbb{Z}.
\label{matching_condition}
\end{equation}

This matching condition can be easily visualized in Fig.\ref{Bloch_structure} for the $N=4$ and $N=5$ cases. When $N=4$ and $l \le 2$ we can always find Bloch states in the first Brillouin zone with the same value of the angular pseudo-momentum $m=l$. This is the case of the input angular momentum $l=2$ represented in Fig.\ref{Bloch_structure}(a). For $N=5$ the situation is similar, i.e., we excite angular Bloch states with $m=l$ as long as $l\le 2$ (see Fig.\ref{Bloch_structure}(b)). However, when this condition is not fulfilled ($l>2$) we excite an equivalent Bloch state with different angular pseudo-momentum in the first Brillouin zone.
Let us choose, for example, the input angular momentum to be $l=3$ in the $N=5$ case. We immediately see that we are exciting an equivalent Bloch state with different angular pseudo-momentum in the first Brillouin zone given by $m=-2$, in agreement with the matching condition
 (\ref{matching_condition}).

The previous considerations refer to angular pseudo-momentum and thus they are related to the way the wavefunction transform under discrete rotations; i.e., in the way described by 
Eq.(\ref{transformation_property}). Using analogous arguments as those appearing in Ref.\cite{9}, it can be proven that the angular pseudo-momentum is conserved during time evolution. This means that if we start at the initial time $t_0$ with a solution fulfilling Eq.(\ref{transformation_property}), this solution will preserve this property during evolution provided no new alteration in the symmetry of the potential is given. In other words, if the initial angular pseudo-momentum is $m$ at $t_0$ it will remain the same at any other time $t>t_0$. This property has been numerically checked in optical vortex transmutation phenomena \cite{13}. When, besides having a solution with angular pseudo-momentum $m$, we are dealing with a function characterized by a single-phase singularity the angular pseudo-momentum $m$ becomes identical to the topological charge \cite{12,13}.

It is remarkable that the irreducible representations in $C_{nv}$ are either one or two dimensional. This statement is based in group theory arguments or can be deduced directly from inspection of the bloch functional form (\ref{bloch_modes}). In fact, if $N$ is even, the dimension of the representation is one for $m=0$ or $\mid m\mid =\frac{N}{2}$, while it is of dimension two for $\mid m\mid =1,\dots,\frac{N}{2}-1$. And if $N$ is odd, the dimension of representation is one for $m=0$, while it is two for $\mid m\mid  =1,\dots,\frac{N-1}{2}$. Since one dimensional representations  do not transform under rotations except for a sign, they are necessarily real. On the other hand, two dimensional representations present a non trivial phase structure. Thereby, only solutions belonging to two dimensional irreducible representations can be considered as vortex solutions. Notice that, as stated previously, the base  for these two dimensional irreducible representations is the complex pair $\psi_m$ and $\psi^\ast_m$ of vortex-antivortex soliton solutions.

According to these statments and to the matching condition (\ref{matching_condition}) there are different kinds of vorticity transformations between $O(2)$ and  $C_{nv}$ media.  For $l$ in the first Brillouin zone no transformation is produced. For $l$ outside the first Brillouin zone there are two kinds of transformation: i) if the value of $m$ that corresponds to the input $l$ according to the matching condition (\ref{matching_condition}) is such that the solution belongs to a irreducible representation of dimension one, the phenomenon is called {\it charge erasing} and ii) if $m$ is such that the related representation is a two dimensional one, the phenomenon is called {\it vortex transmutation}. Moreover, there are two kinds of vortex transmutation phenomenon: i) charge downconversion if $m$ has the same sign that $l$ and ii) charge inversion if $m$ has different sign that $l$.  For example, in in Fig.\ref{Bloch_structure}(a) a case of charge erasing is presented. On the other hand, the Fig.\ref{Bloch_structure}(b) represents a charge inversion.

This remarkable fact can be used to exploit the angular pseudo-momentum matching condition (\ref{matching_condition}) as a way to control the topological charge of matter wave vortices, as it will demonstrated extensively in the next sections.

\section{Applications}
\label{App}

\subsection{Practical configuration}
\label{Pra}

In order to work with the minimal configuration showing the phenomena to be described here we have made the simplifications of
taking the linear limit $g=0$ and chosing a simple potential $V(x,y)$ obtained as the superposition of $N$ gaussian functions of the form
\begin{equation}\label{pot}
V(x,y) = V_0 \sum_{j=0}^{N-1} \exp\left\{ \left[(x-x_j)^2+(y-y_j)^2\right]/(2 w^2)\right\},
\end{equation}
with $(x_j,y_j) = d\left(\cos(2\pi j/N),\sin(2\pi j/N)\right)$
This type of potential can be obtained physically by using a set of laser beams generating a set of optical dipole traps for the Bose-Einstein condensate.

We have checked that our results are essentially independent of the nonlinearity and that they remain valid for the more complicated case of a periodic potential of the required symmetry. However, the choice of the potential as given by Eq. (\ref{pot}) is not only simpler but allows for more freedom in the selection of the symmetry since we can take e.g. $N=5$ to obtain a discrete symmetry of fifth order, something which is not possible with lattice potentials.

\begin{figure}
\epsfig{file=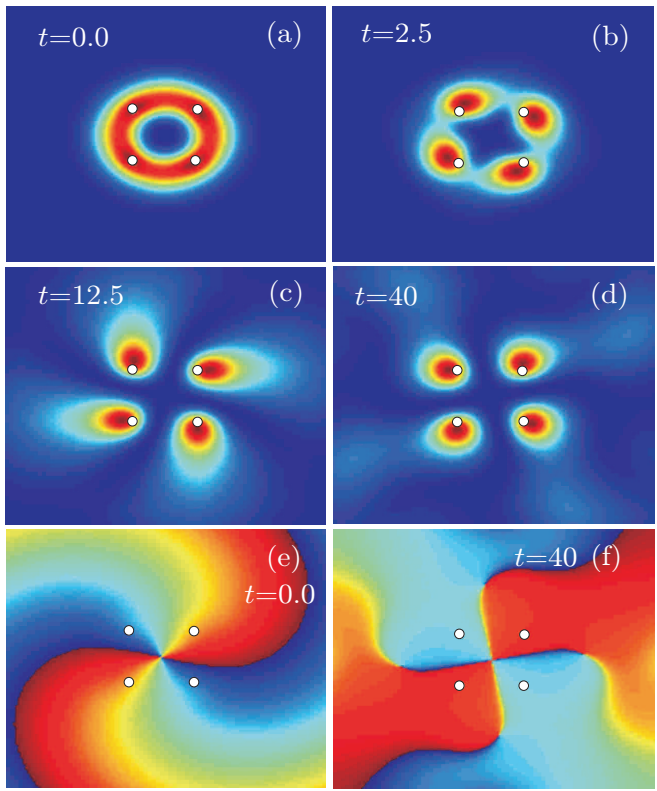,width=\columnwidth}
\caption{[Color online] Erasing of a vortex of topological charge $m=2$ due to the effect of the discrete symmetry of the potential \eqref{pot} with $N=4, d=4, V_0 = -0.6$.  Shown are pseudocolor plots of the amplitude $|\psi(x,y,t)|^2$ (a,b,c,d) and phase $\arg(\psi(x,y,t))$ (e,f) for different times indicated on the subplots. The locations of the potential minima according to Eq. (\ref{pot}) are indicated with small white circles. \label{sec}}
\end{figure}

It is also very simple to change between configurations with different symmetry orders ($N$) even dynamically by
just adding or eliminating  laser beams.

In what follows we will solve Eq. (\ref{NLS}) with $V(x,y)$ given by Eq. (\ref{pot}) and initial data of (multicharged) vortex type.
All simulations to be presented in this paper have been done using a split-step method where the spatial derivatives are computed by using a pseudoespectral formula
based on trigonometric polynomials. In all cases to be presented here we have chosen $\Delta t = 0.025$ and the simulation region $[-20,20]\times [-20,20]$. In the figures the spatial region shown is
$[-10,10]\times [-10,10]$. The outgoing radiation is elliminated by absorbing boundary conditions such as the ones implemented in Ref. \cite{IMACS}.

\subsection{Charge erasing}
\label{erase}

As a first example of vorticity control by discrete symmetries we consider the evolution of an initial configuration of the form
\begin{equation}\label{doubly}
\psi(x,y) = (x+iy)^2 \exp{\left[(-x^2-y^2)/8\right]}
\end{equation}
i.e. a vortex of topological charge $l=2$. This initial wavefunction
will be subject to a potential with fourth order symmetry corresponding to Eq. (\ref{pot}) with $N=4$. The other potential parameters are taken to be $V_0 = -0.6$ and $
d=4$.

 The evolution of this initial datum is shown in Fig. \ref{sec}.  Where some typical features are seen. First of all the amplitude of the wavefunction [Fig. \ref{sec}(a,b,c,d)] becomes localized on
 the four potential wells after a transient in which some radiation is emitted. To achieve this effect of localization it is convenient to adjust the potential minima to be close to the
 initial maximum density so that the fraction of the initial number of particles which is retained into the system is maximized.

At the same time the phase [Fig. \ref{sec}(e,f)] experiences a complicated evolution from the initial configuration with $l=2$, to a final configuration with two nodal lines with $x=0$ and $y=0$ (which are the symmetry axes of this potential) and constant phases across the four regions in which the space is separated by them, which correspond to a solution with $m=2$ that belongs to a representation of dimension one.  Let us obtain the theoretical prediction for this case. We must take $k=0$ in expression (\ref{matching_condition}),  since the input vortex is in the first Brillouin zone, and consequently, $m=2-0\cdot 4=2$, in agreement with the numerical result.

\subsection{Generation of singly charged vortices from doubly charged ones: charge inversion}
\label{single}

\begin{figure}
\epsfig{file=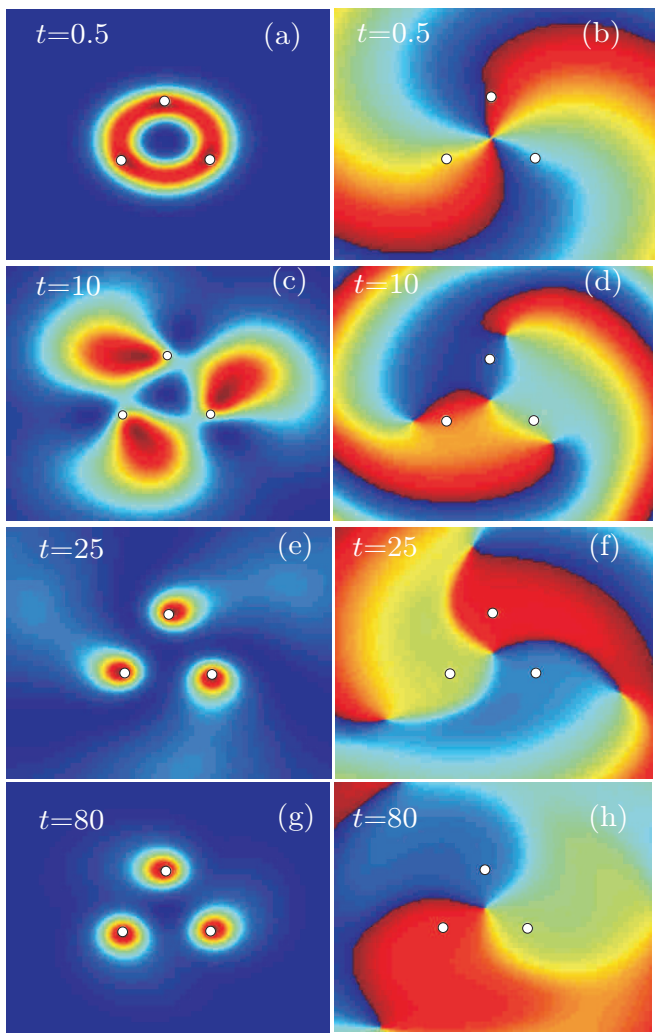,width=\columnwidth}
\caption{[Color online] Vorticity inversion due to the effect of the discrete symmetry of the potential \eqref{pot} with $N=3, d=3.2, V_0 = -0.6$.  Shown are pseudocolor plots of the amplitude $|\psi(x,y,t)|^2$ (a,c,e,g) and phase $\arg(\psi(x,y,t))$ (b,d,f,h) for different times indicated on the subplots. The locations of the potential minima according to Eq. (\ref{pot}) are indicated with small white circles. \label{sym3}}
\end{figure}

An interesting problem is how to process the doubly charged vortices obtained in atom chips by phase imprinting methods \cite{Kett1,Kett2} to obtain vortices with charge $m=-1$.

As in the previous examples, discrete symmetries can be helpful to accomplish this task. Let us take now a potential with $N=3$ and a doubly charged vortex as initial datum as in Eq. (\ref{doubly}).
Let  us choose the potential parameters as $d=3.2$, and $V_0 = -0.6$.

 The evolution of this configuration is shown in Fig. \ref{sym3} where the inversion of the central vortex from $l=2$ to $m=-1$ is clearly seen in the phase plots [Fig. \ref{sym3}(b,d,f,h)]. In this case, to obtain the theoretical prediction, we must take $k=1$, since $l=2$ falls in the second Brillouin zone for $N=3$. Consequently, according to the matching condition (\ref{matching_condition}), $m=2-1\cdot 3=-1$.
It can be seen how initially a central vortex with charge $m=-1$ coexists with other vortices located around it. However, as time becomes larger these vortices slowly spiral out of the system and are not visible in the region of interest for $t \sim 80$.

The transmutation phenomenon has been studied previously in Refs. \cite{13,Ripi}. However, in Ref. \cite{Ripi} the asymmetric trap used to induce the phenomenon leads to a recurrence in which the topological charge periodically oscillates between +1 and -1, while in this setup we get $m=-1$ for any time larger than the transient time in which the system becomes stabilized. With respect to the proposal of Ref. \cite{13} the scheme presented here has the advantage of introducing less radiation and being cleaner because of the finite range of the potential.

\subsection{Obtention of a vortex of charge $m=-1$ from a vortex with charge $l=4$ in a system with symmetry $N=5$: charge inversion}
\label{minusingle}

\begin{figure}
\epsfig{file=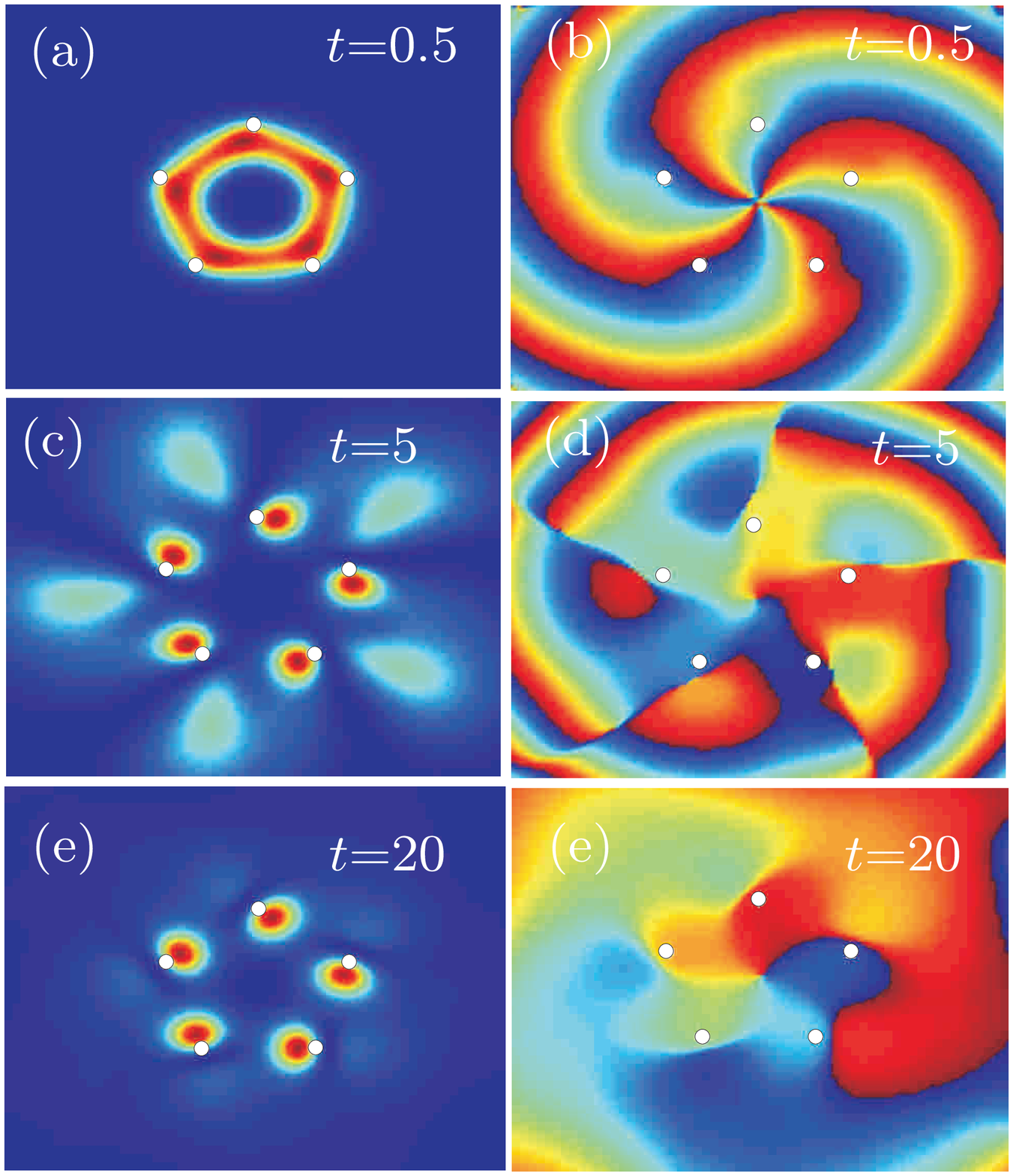,width=\columnwidth}
\caption{[Color online] Vorticity inversion from $m=4$ to $m=-1$ due to the effect of the discrete symmetry of the potential \eqref{pot} with $N=5, d=3.2, V_0 = -2$.  Shown are pseudocolor plots of the amplitude $|\psi(x,y,t)|^2$ (a,c,e) and phase $\arg(\psi(x,y,t))$ (b,d,f) for different times indicated on the subplots. The locations of the potential minima according to Eq. (\ref{pot}) are indicated with small white circles. \label{sym5}}
\end{figure}

As a final example we show how to obtain a vortex with charge $m=-1$ from a vortex with topological charge $l=4$. To do so we must use a system with symmetry $N=5$.
Our initial wavefunction is given by
\begin{equation}\label{four}
\psi(x,y) = (x+iy)^4 \exp{\left[(-x^2-y^2)/2\right]},
\end{equation}
and  the parameters in the potential are taken as  $V_0 = -2$ and $d=3.2$.

As in the previous examples we observe emission of ratiation (in this case stronger due to the increasing energy of the initial condition which leads to a faster scape from the origin) and after some time, the desired structure is observed. Again, $k=1$ since $l=4$ falls in the second Brillouin zone for $N=5$. Thereby, according to (\ref{matching_condition}), $m=4-1\cdot 5=-1$, in agreement with the numerical solution.

\section{Conclusions and discussion}
\label{Con}

In this paper we have studied the dynamics of multiply-charged vortices under the action of potentials with discrete symmetries. We have shown how the symmetry order of the potential can be used to manipulate the topological charge of the vortex in order to obtain a desired (lower) value of the vorticity. Specifically we have proposed a setup of gaussian traps which can be configured to have the appropriate symmetry and more flexibility than the lattice type potentials previously used.

As applications we have studied: topological charge erasing, charge inversion from $l=2$ to $m=-1$ and topological charge conversion from $l=4$ to $m=-1$
The last two examples are applicable to the controlled generation of  singly charged vortices from the output of the phase imprinting method using matter-wave chips developed in Ref. \cite{Kett1}.

The vortex charge control and erasing properties which can be achieved in this system open new possibilites for the control of quantum matter. Moreover, although we have choosen an specific model given by Eqs. (\ref{NLS}) and (\ref{pot}) our arguments developed in Sec. \ref{III} are completely general and depend only on the discrete symmetry properties of the system. Thus, they can be extended beyond the specific form of the potential choosen and even the Gross-Pitaevskii equations used in this paper.

\acknowledgments

This work has been partially supported by grants BFM2003-02832, FIS2004-20188-E, FIS2005-01189, and FIS2006-04190  (Ministerio de Educaci\'on y Ciencia, {S\-pa\-in}),  PAI05-001 (Consejer\'{\i}a de Educaci\'on y Ciencia de la Junta de Comunidades de Castilla-La Mancha).

\end{document}